\documentclass[a4paper,11pt]{article}
\pdfoutput=1 

\usepackage{aasmacros}
\usepackage{jcappub} 

\usepackage[T1]{fontenc} 

\usepackage{natbib}
\usepackage{graphicx}
\usepackage{graphics,color}
\usepackage{soul, xcolor}
\usepackage[normalem]{ulem}
\usepackage{float}
\usepackage{hyperref}
\usepackage{booktabs}
\hypersetup{
    colorlinks=true,
    linkcolor=blue,
    filecolor=magenta,      
    urlcolor=cyan,
}

\usepackage{subfigure}

\newcommand{\ba}{\begin{eqnarray}}
\newcommand{\ea}{\end{eqnarray}}

\newcommand  \beq    {\begin{equation}}

\newcommand  \eeq    {\end{equation}}
\newcommand  \gtsim  {\lower.5ex\hbox{$\; \buildrel > \over \sim \;$}} 
\newcommand  \ltsim  {\lower.5ex\hbox{$\; \buildrel < \over \sim \;$}}

\graphicspath{{./figs/}}

\definecolor{so-orange}{RGB}{242, 101, 35}

\newcommand{\numbered}[2]{#2}

\newcommand{\crossout}[1]{}

\emailAdd{zequnl@astro.princeton.edu}

\title{The cross correlation of the ABS and ACT maps.  }

\author{Zack Li$^1$}
\author{Sigurd Naess$^2$}

\author{Simone Aiola$^{2}$}
\author{David Alonso$^{25}$}
\author{John W. Appel$^{10}$}
\author{J. Richard Bond$^{15}$}
\author{Erminia Calabrese$^{11}$}
\author{Steve K. Choi$^{14}$}
\author{Kevin T. Crowley$^{21}$}
\author{Thomas Essinger-Hileman$^{10}$}
\author{Shannon M. Duff$^{5}$}
\author{Joanna Dunkley$^{1,3}$}
\author{J. W. Fowler$^{5,6}$}
\author{Patricio Gallardo$^{13}$}
\author{Shuay-Pwu Patty Ho$^{3}$}
\author{Johannes Hubmayr$^{5}$}
\author{Akito Kusaka$^{26, 27, 28}$}
\author{Thibaut Louis$^{19}$}
\author{Mathew S. Madhavacheril$^{22}$}
\author{Jeffrey McMahon$^{29, 30, 31}$}
\author{Federico Nati$^{20}$}
\author{Michael D. Niemack$^{13,14}$}
\author{Lyman Page$^{3}$}
\author{Lucas Parker$^{17}$}
\author{Bruce Partridge$^{16}$}
\author{Maria Salatino$^{8,9}$}
\author{Jonathan L. Sievers$^{23,24}$}
\author{Crist\'obal Sif\'on$^7$}
\author{Sara M. Simon$^{12}$}
\author{Suzanne T. Staggs$^{3}$}
\author{Emilie Storer$^{3}$}
\author{Edward J. Wollack$^{18}$}

\affiliation{$^1$Department of Astrophysical Sciences, Princeton University, 4 Ivy Lane, Princeton, NJ, USA 08544}
\affiliation{$^2$Center for Computational Astrophysics, Flatiron Institute, 162 5th Avenue, New York, NY, USA 10010}
\affiliation{$^3$Joseph Henry Laboratories of Physics, Jadwin Hall, Princeton University, Princeton, NJ 08544, USA}
\affiliation{$^4$Department of Astronomy and Astrophysics, The Pennsylvania State University, University Park, PA 16802, U.S.A.}
\affiliation{$^5$National Institute of Standards and Technology, Boulder, Colorado 80305, USA}
\affiliation{$^6$Department of Physics, University of Colorado, Boulder, Colorado 80309, USA}
\affiliation{$^7$Instituto de F\'isica, Pontificia Universidad Cat\'olica de Valpara\'iso, Casilla 4059, Valpara\'iso, Chile}
\affiliation{$^8$Department of Physics, Stanford University, Stanford, CA 94305}
\affiliation{$^9$Kavli Institute for Particle Astrophysics and Cosmology, Menlo Park, CA 94025, U.S.A.}
\affiliation{$^{10}$Department of Physics and Astronomy, The Johns Hopkins University, Baltimore, MD 21218-2686, USA}
\affiliation{$^{11}$School of Physics and Astronomy, Cardiff University, The Parade, Cardiff, CF24 3AA, U.K. }
\affiliation{$^{12}$Department of Physics, University of Michigan, Ann Arbor, 48103, U.S.A.}
\affiliation{$^{13}$Department of Physics, Cornell University, Ithaca, NY 14853, U.S.A.}
\affiliation{$^{14}$Department of Astronomy, Cornell University, Ithaca, NY 14853, U.S.A.}
\affiliation{$^{15}$Canadian Institute for Theoretical Astrophysics, University of Toronto, 60 St. George St., Toronto, ON M5S 3H8, Canada}
\affiliation{$^{16}$Dept. of Astronomy, Haverford College, Haverford PA 19041, USA}
\affiliation{$^{17}$Space and Remote Sensing, MS D436, Los Alamos National Laboratory,
Los Alamos, NM 87544, USA}
\affiliation{$^{18}$NASA/Goddard Space Flight Center, Greenbelt, MD 20771, USA}
\affiliation{$^{19}$LAL, Univ. Paris-Sud, CNRS/IN2P3, Universit\'e Paris-Saclay, Orsay, France}
\affiliation{$^{20}$Department of Physics, University of Milano - Bicocca, Piazza della Scienza 3, I-20126 Milano, Italy}
\affiliation{$^{21}$Department of Physics, University of California, Berkeley, CA, USA 94720}
\affiliation{$^{22}$Centre for the Universe, Perimeter Institute for Theoretical Physics, Waterloo, ON, Canada N2L 2Y5}
\affiliation{$^{23}$Department of Physics, McGill University, 3600 Rue University, Montreal, Quebec H3A 2T8, Canada}
\affiliation{$^{24}$School of Chemistry and Physics, University of KwaZulu-Natal, Private Bag x54002 Durban, South Africa}
\affiliation{$^{25}$Department of Physics, University of Oxford, Denys Wilkinson Building, Keble Road, Oxford OX1 3RH, United Kingdom}
\affiliation{$^{26}$Physics Division, Lawrence Berkeley National Laboratory, Berkeley, CA 94720, USA}
\affiliation{$^{27}$Kavli Institute for the Physics and Mathematics of the Universe,  5-1-5 Kashiwanoha,
Kashiwa 277-8583, Japan}
\affiliation{$^{28}$Department of Physics, University of Tokyo, Tokyo 113-0033, Japan}
\affiliation{$^{29}$Kavli Institute for Cosmological Physics, University of Chicago, 5640 S. Ellis Ave., Chicago, IL 60637, USA }

\affiliation{$^{30}$Department of Astronomy and Astrophysics, University of Chicago, 5640 S. Ellis Ave., Chicago, IL 60637, USA} 

\affiliation{$^{31}$Department of Physics, University of Chicago, Chicago, IL 60637, USA}

\abstract{
One of the most important checks for systematic errors in CMB studies is the cross correlation of maps made by independent experiments. In this paper we report on the cross correlation between maps from the Atacama B-mode Search (ABS) and Atacama Cosmology Telescope (ACT) experiments in both temperature and polarization. These completely different measurements have a clear correlation with each other and with the Planck satellite in both the EE and TE spectra at $\ell<400$ over the roughly $1100$ deg$^2$ common to all three. The TB, EB, and BB cross spectra are consistent with noise. Exploiting such cross-correlations will be important for future experiments operating in Chile that aim to probe the $30<\ell<8,000$ range.
}

\begin{document}

\maketitle
\flushbottom

\section{Introduction}
\label{sec:intro}

The primary data products of most CMB experiments are temperature and polarization maps. From these, one may compute power spectra, model foreground emission, compute the \crossout{the} lensing map, and investigate many other phenomena. Maps have the advantage that with them the results of different experiments can be compared directly. This is crucial for assessing systematic errors and will become increasingly important as the noise per pixel drops well below that achieved by the Planck satellite. While many CMB experiments are calibrated to WMAP\citep{bennett/etal:2013} or Planck\citep{planck_overview:2016}, few ground or balloon-based experiments have been compared to each other. Since 2000, to our knowledge the Atacama Cosmology Telescope (ACT) and South Pole Telescope (SPT) are the only independent ground-based maps which have been compared directly \citep{dunner/etal:2013}. In this paper, we show that maps from the Atacama B-mode Search (ABS) and ACT are correlated in polarization, agree in the polarization-temperature correlation, and agree with expectations from $\Lambda$CDM.

There is much more to learn from the CMB 
at both large and small angular scales (\numbered{11 (remove colons)}{}  e.g., Simons Observatory \cite{SO:2019},
CMB-S4 \cite{CMB-S4:2019}, LiteBird \numbered{11}{\crossout{\cite{litebird:2018}} \cite{litebird/etal:2020}}, PICO \cite{hanany/etal:2019}). From the ground and balloons, a range of angular scales are being covered as currently exemplified by BICEP \citep{bicep/keck:2018}, CLASS \citep{essinger-hileman/etal:2014}, Polarbear/Simons Array \citep{pbear-BB:2019}, SPIDER \citep{spider/etal:2008},
ACT \citep{louis/etal:2017}, and SPT \citep{sayre/etal:2019}. The ability to cross-correlate these experiments will enhance the results from each, and be an important check for systematic effects. In this paper, we cross-correlate polarization maps from ABS, an experiment targeting large scales ($\ell < 400$), with ACT, an experiment which principally targets the small scales ($\ell > 350$).

\begin{figure}
    \centering
    \includegraphics[width=0.45\textwidth]{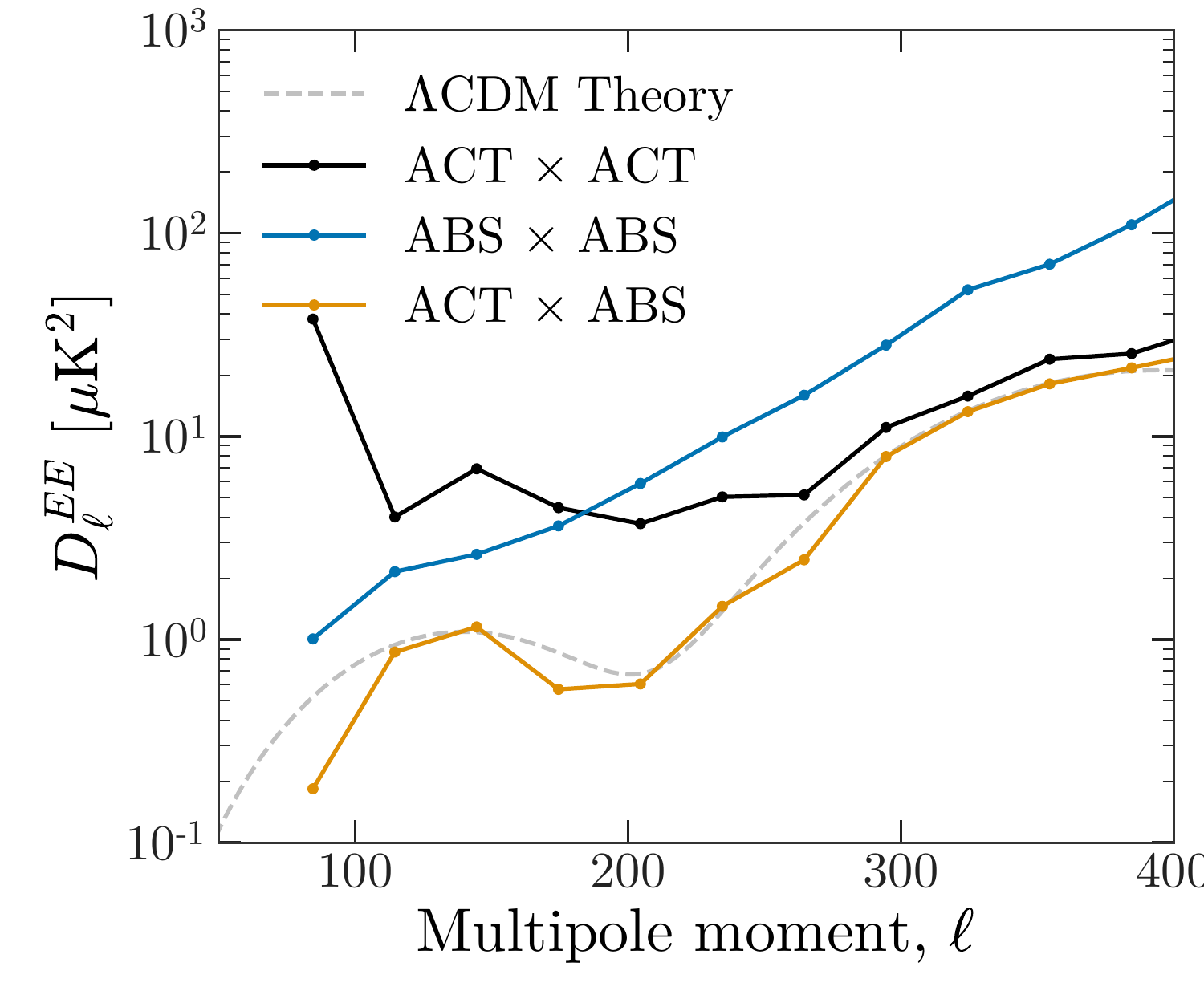}
    \caption{EE auto- and cross-spectra between polarization maps for ACT, ABS, demonstrating the noise properties of the different experiments. The auto spectra include the noise bias. A theoretical prediction for the spectrum from a $\Lambda$CDM model is shown in dashed gray, with parameters given in $\S$\ref{sec:analysis}.}
    \label{fig:autos_and_crosses}
\end{figure}

\section{Observations}
\label{sec:obs}

\subsection{ABS}
The ABS experiment is described in \cite{simon/etal:2014}, \cite{essinger-hileman/etal:2016}, and \cite{kusaka/etal:2018}. The instrument has a $0.59\,$m diameter $4\,$K crossed Dragone telescope that feeds a 
roughly hexagonal array of 240 horn-coupled TES dual-polarization bolometers fabricated at NIST \citep{bleem/etal:2009} cooled to 0.3K and measuring in a band centered near 145 GHz. The TESs are read out with the time-division multiplexing electronics \citep{battistelli/etal:2008}.  A rotating ($f_{rot}=2.55\,$Hz) ambient-temperature half-wave plate (HWP) is the first optical element, and sits just outside the vacuum window near the aperture stop. The focal plane maps to a field of view 22$^\circ$ across on the sky. The average beam in the array has a full-width-at-half maximum of  $32^\prime$.

To observe, the optical axis is set at a $45^\circ$ elevation angle and the instrument scans back and forth at approximately constant speed, with a 27 s period and amplitude in azimuth of $10^\circ$. The HWP modulates the incident polarized signal at $4f_{\text{rot}}=10.2\,$Hz as measured at the detector output. This is well above the 1-2 Hz $1/f$ knee of the atmosphere for these detectors which typically have a white noise level of 580 $\mu$Ks$^{1/2}$ (NET). The low-$\ell$ limit of the power spectra is set by the map dimensions rather than by atmospheric fluctuations. Data were taken between September 2012 and December 2013 in ABS ``field A'' which is centered on RA$=25^\circ$, Dec$=-41^\circ$.

The incident signal is demodulated as described in \cite{kusaka/etal:2014}. The demodulated data are binned using the Healpix package \citep{gorski/etal:2005} with $N_{\text{side}}=256$. The ABS Stokes $Q$ and $U$ are defined relative to equatorial coordinates.\footnote{The ABS, ACT, and Planck polarization maps are made in Stokes $Q$ and $U$ parameters following the Healpix convention \citep{gorski/etal:2005}. To find a polarization angle that follows the IAU convention \citep{hamaker/bregman:1996} compute 
$\gamma_P=(1/2){\rm arg}(Q-iU)$.} The $Q$ and $U$ maps, beam window function $W_\ell$, inverse-variance weighted spatial window (or mask) $M(\vec{x})$, and map transfer function are all available on LAMBDA\footnote{LAMBDA: https://lambda.gsfc.nasa.gov/product/abs/index.cfm }. The raw maps cover 2400 deg$^2$. The area inside the roughly half power point (0.44) of the spatial window maximum is 1126 deg$^2$ ($f_{sky}=0.027$). A typical noise level in the $Q$ and $U$ maps is $4.5~\mu$K with respect to the CMB in an $N_{side}=256$ pixel (approximately $13.8^\prime\times 13.8^\prime$). The calibration uncertainty (including the beam uncertainty) in power ranges from 15\% at $\ell\approx50$ to 12\% at $\ell\approx350$. See \cite{kusaka/etal:2018} for more details.

The ABS maps are not maximum-likelihood estimates. Multiple end-to-end simulations of the time streams passed through the full data-analysis pipeline were generated to compute the transfer functions as well as the power spectra and uncertainties.

\subsection{ACT}
The ACT experiment is described in \cite{thornton/etal:2016}. The receiver is in its third generation called Advanced ACTPol (AdvACT, \cite{henderson/etal:2016}, \cite{simon/etal:2016}, \cite{choi/etal:2018}, \cite{crowley/etal:2018}). ACT has been observing since 2007. The region covered by ABS was observed in 2016 through 2018.\footnote{ACT adopted a blinding strategy in the fourth data release (DR4) to mitigate confirmation bias. That data set is independent of the data presented in this paper and focuses on higher $\ell$.} 

AdvACT has three separate arrays of dichroic dual polarization horn-coupled TES bolometers fabricated at NIST, cooled to 0.1K and measuring in bands near 90~GHz, 150~GHz, and 220~GHz \citep{li/etal:2016,choi/etal:2018,ho/etal:2016}. Each array is housed in a separate ``optics tube'' that also holds the filters \citep{tucker/ade:2006} and optical elements.
In this paper we consider only 90 and 150~GHz. As with ABS, the TESs are read out with time-division multiplexing electronics.  The field of view of one optics tube is about $1^\circ$. The average beam in the array has a full-width-at-half 
maximum of $1.4^\prime$ at 150 GHz and $2.1^\prime$ at 90 GHz.

The maps used for this paper were made in two steps. The first step uses a maximum likelihood code to make individual maps
for each observing season, frequency and detector set. Descriptions of the pipeline are given in \cite{dunner/etal:2013},
\cite{naess/etal:2014}, and \cite{louis/etal:2017}. The native maps are produced in the Plate Carr\'ee (CAR) projection with $0.5^\prime$ pixels. The second step combines these maps into a single map per frequency,
reconvolving to a common beam and weighting each map using a noise model consisting of a hitcount-modulated 2d noise power spectrum
for each $4^\circ\times 4^\circ$ tile of the maps (Naess et al. in preparation, see also \cite{chown/etal:2018}). The 90 GHz and 150 GHz maps are then coadded after convolving to a common beam.  One of the benefits of maximum-likelihood maps is that
they can be nearly unbiased - the transfer function can be close to unity to low $\ell$. 
Because the map-making process for the preliminary 2017-2018 ACT maps
used here was not run to full convergence especially in temperature\footnote{ACT
maximum-likelihood maps are solved using conjugate gradient (CG) iteration.
We generally use three phases of 300 iterations for this, with the noise model
being re-estimated after each phase, for a total of 900 CG steps. The
preliminary maps, because of their size, were made with a single phase of 300 iterations.},
these particular maps do have a significant bias at low $\ell$ (as demonstrated in Figure~\ref{fig:transfer}), which we correct as discussed
below.


The resulting map covers 18,300 deg$^2$ in 380 Megapixels (the full map including unfilled pixels is 415 Mpix or 20,000 deg$^2$). The maps follow the Healpix polarization convention defined relative to equatorial coordinates and has a calibration uncertainty of 5\%, as determined by comparison to Planck temperature maps. The maps used for this paper have not yet undergone the battery of
tests necessary for public release and are thus preliminary.
However, based on our tests the maps are sufficient for the analysis presented here, given the relative immunity of cross-correlations to experiment-specific systematic effects. 
These maps are rebinned into the Healpix format in the native resolution and downgraded to $N_{side}=256$ for the cross correlation with ABS. 

Figure~\ref{fig:autos_and_crosses} shows the EE auto and cross spectra for ABS and ACT. 
In a typical ACT pixel at $N_{side}=256$  in the ABS region, the noise in the $Q$ and $U$ maps is $1.1~\mu$K relative to the CMB in the combined
at 90 GHz and 150 GHz maps. 
While ABS has higher white (detector) noise than ACT, the noise fluctuations at large scales are smaller because of the HWP demodulation. By comparison,
the noise in the Planck $Q$ and $U$ maps at the same resolution in this region is typically 4.4 $\mu$K at 143 GHz, comparable to the ABS noise.

\section{Analysis and discussion}
\label{sec:analysis}

\begin{figure}
    \centering
    \includegraphics[width=0.45\textwidth]{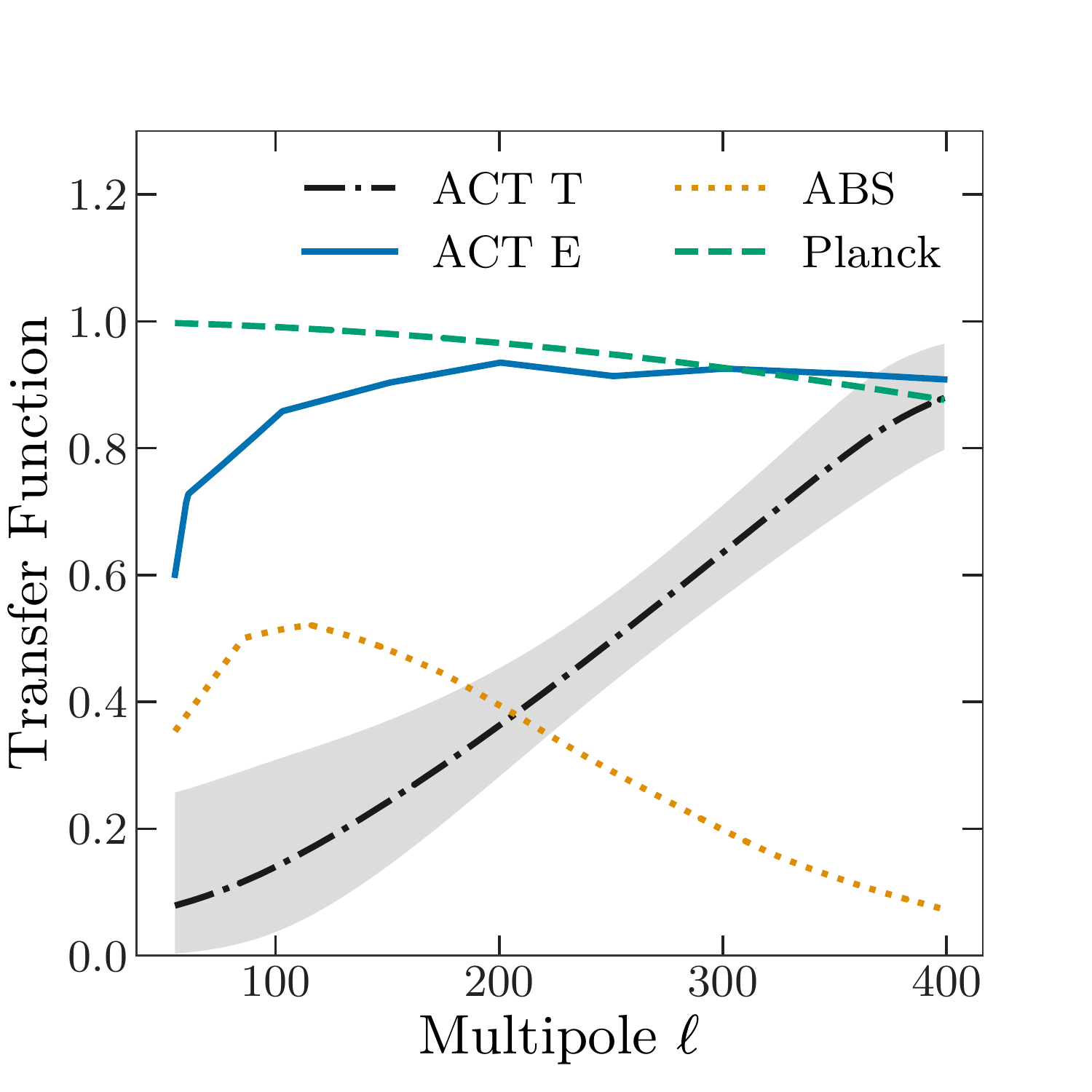}
    \caption{
    \numbered{1 (added error bar to ACT T figure)}{}
    Products of the window and transfer functions in power. The ACT transfer functions are determined through cross correlations with Planck that are independent of the comparison to ABS, as described in the text. These transfer functions come from unconverged maps which are not typical for ACT analysis, and we expect the transfer function to be closer to unity upon further map processing in future analyses. The ABS transfer function is estimated from simulations and reported in \cite{kusaka/etal:2018}.}
    \label{fig:transfer}
\end{figure}

\begin{figure*}
    \centering
    \includegraphics[width=\textwidth]{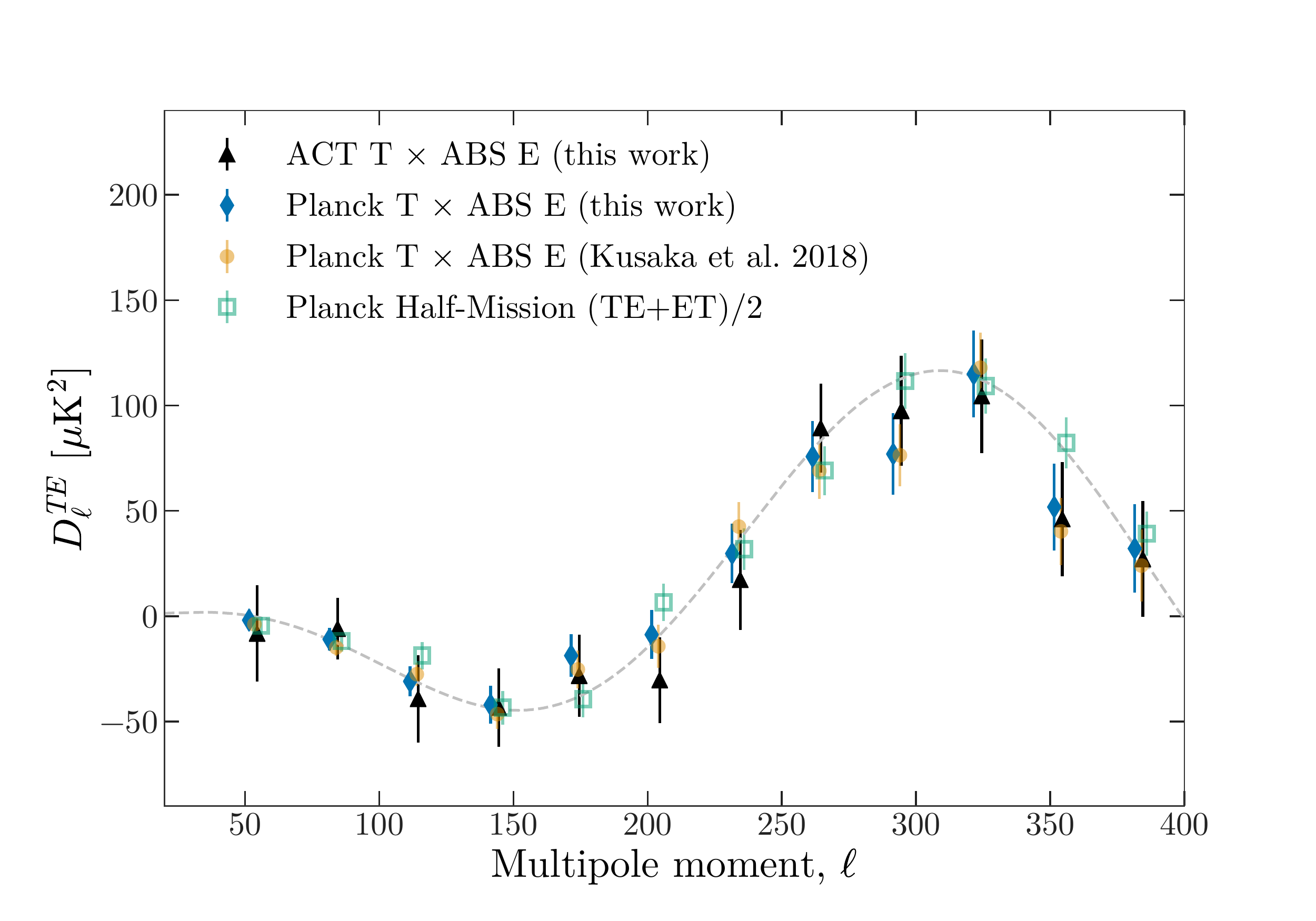}
    \caption{Cross-spectra between temperature and polarization maps for ACT, ABS, and Planck experiments restricted to the ABS patch. The similarity of the blue and orange points indicates that the pipeline for this paper independently reproduces the results in \cite{kusaka/etal:2018}. Table~\ref{tab:chisq} shows the $\chi^2$ between the black triangles (ACT $T$ $\times$ ABS $E$) and green squares (Planck half-mission spectra).  A theoretical prediction for the spectrum from a $\Lambda$CDM model is shown in dashed gray, with parameters given in $\S$\ref{sec:analysis}. Data points at a specific $\ell$ are offset for clarity. \numbered{1}{We do not account for the error in estimated transfer function in ACT, as we expect the function to be highly correlated bin-to-bin.}}
    \label{fig:TE}
\end{figure*}

The inputs for the analysis are the ABS $Q$ and $U$ maps, ABS simulations and transfer function, the ACT $Q$, $U$ and $T$ maps, and the Planck half-mission 
$Q$, $U$, and $T$ maps at 143 GHz. All maps are converted to $N_{side}=256$ 
by transforming to spherical harmonics and then reprojecting. The maps are masked to $\sim 1100$ deg$^2$ with the ABS mask. 
The cross spectra are formed between combinations of $E$ and $B$ polarization for all three experiments, and between Planck and ACT $T$. \numbered{4}{For cross-spectra between Planck and other experiments, we use the sum of the half-mission maps. We combine the masks by computing the minimum of the half-mission masks.}

For comparison to theory, we use spectra from the Boltzmann code CLASS \citep{blas/etal:2011} with $\Lambda$CDM parameters consistent with Planck 2015 \citep{planck_I:2015}, $\omega_b = 0.022032$, $\omega_{cdm} = 0.12038$, $h=0.67556$. $A_s = 2.215 \times 10^{-9}$, $n_s = 0.9619$.

The analysis uses the general-purpose CMB power-spectrum code \texttt{nawrapper}\footnote{Available through Github at \href{https://github.com/xzackli/nawrapper}{xzackli/nawrapper}} based on the NaMaster pseudo-$C_{\ell}$ code \citep{alonso/etal:2019}. The \texttt{nawrapper} code is a component of the Simons Observatory power spectrum pipeline, and has been validated with both independent internal codes on simulations, as well as on external datasets.
When applied to the publicly available Planck half-mission maps, the code reproduces the Planck \numbered{4}{TT, TE, and EE power spectra analyzed using the Plik, the primary pipeline in the 2018 Planck release,} to $0.1 \sigma$ accuracy. Here we adopt the ABS binning for the power spectra, account for the Healpix pixel transfer function, and apply the ABS transfer function derived from simulations. We assume Gaussian covariance matrices when computing the errors, but the noise power spectra are estimated differently depending on the data source. The Planck noise power spectrum is estimated by computing spectrum of the half-mission difference map \numbered{5}{and multiplying by $1/2$.} For ABS
we estimate the noise power spectrum using the mean power spectrum of the ABS noise simulations. 

Although ACT map sets typically include four data splits, these preliminary ACT maps do not have any splits. We therefore estimate the noise in each map by computing the auto-spectrum and subtracting from it a theoretical prediction derived from Planck $\Lambda$CDM parameters. The assumption of $\Lambda$CDM theory spectra affects the products of the covariance matrices such as error bars and null tests, but does not impact the power spectra themselves. In addition, we account for the calibration uncertainty in the covariance matrix diagonals (e.g. \cite{ganga/etal:1997}).

%
%

The ACT transfer functions for the large maps are currently difficult to simulate. We estimate them here by assuming Planck is a true representation of the sky, after accounting for the beam, and taking the ratio of  (ACT $T$)$\times$(Planck $E$)/(Planck $T$)$\times$(Planck $E$) and 
(ACT $E$)$\times$(Planck $E$)/(Planck $E$)$\times$(Planck $E$). The ratios of these cross-spectra are noisy but sample a transfer function which is fairly smooth, thus we fit a polynomial to the ratio. The resulting transfer functions, multiplied by the appropriate beam functions, are shown in Figure~\ref{fig:transfer} along with the same products for Planck and ABS. \numbered{1}{We provide an estimate in Figure~\ref{fig:transfer} of the 1$\sigma$ errors on the ACT transfer function in each bin for TT by performing our same analysis over 100 Gaussian map realizations of the involved spectra using the same mask. The corresponding transfer function in polarization is well-determined from simulations.} We then divide all other ACT spherical harmonic amplitudes by the transfer function at the appropriate $\ell$.  All other cross spectra involving ACT or Planck are independent of these two.\footnote{A cross correlation requires both phases and amplitudes of a signal to match in the maps.} Figure~\ref{fig:transfer} shows that the polarization maps converge more quickly than the temperature maps. This is expected from our iterative mapmaking, as the large-scale noise in temperature from the atmosphere, slow variations in the instrument, and large-scale pickup delays the conjugate gradient solver from solving for the large-scale modes. 

\begin{figure*}
    \centering
    \includegraphics[width=\textwidth]{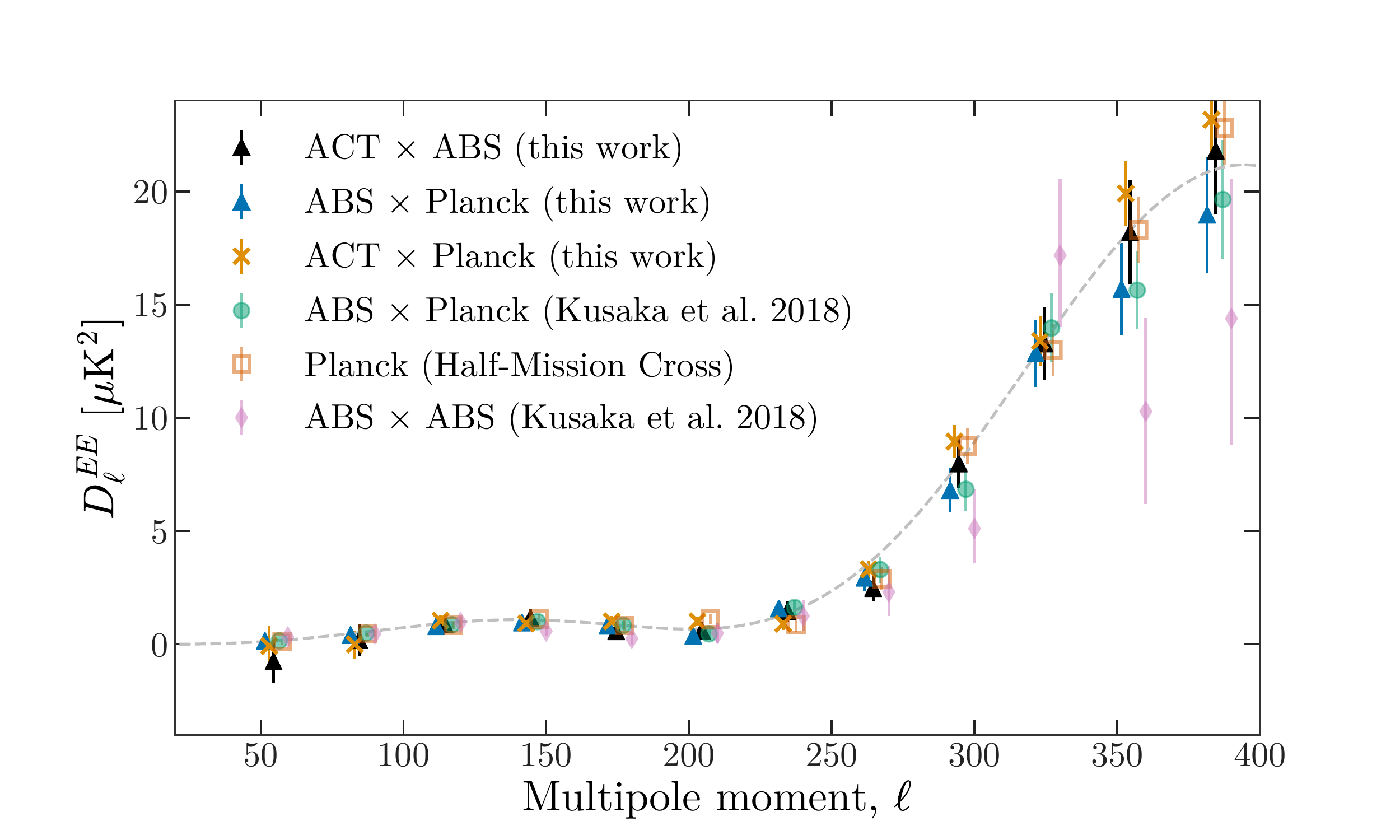}
    \caption{Polarization cross-spectra between ACT, ABS, and Planck restricted to ABS field A. The ACT$\times$Planck spectrum uses ACT E and Planck T and is thus independent of the correlation used to determine the transfer function. A theoretical prediction for the spectrum from a $\Lambda$CDM model is shown in dashed gray, with parameters given in $\S$\ref{sec:analysis}. Note that the ABS$\times$ACT uncertainties are smaller than those of ABS$\times$ABS for $\ell>200$. Data points at a specific $\ell$ are offset for clarity. \numbered{1}{We do not account for the error in estimated transfer function in ACT, as we expect the function to be highly correlated bin-to-bin.}}
    \label{fig:EE}
\end{figure*}

\begin{figure*}
    \centering
    \includegraphics[width=\textwidth]{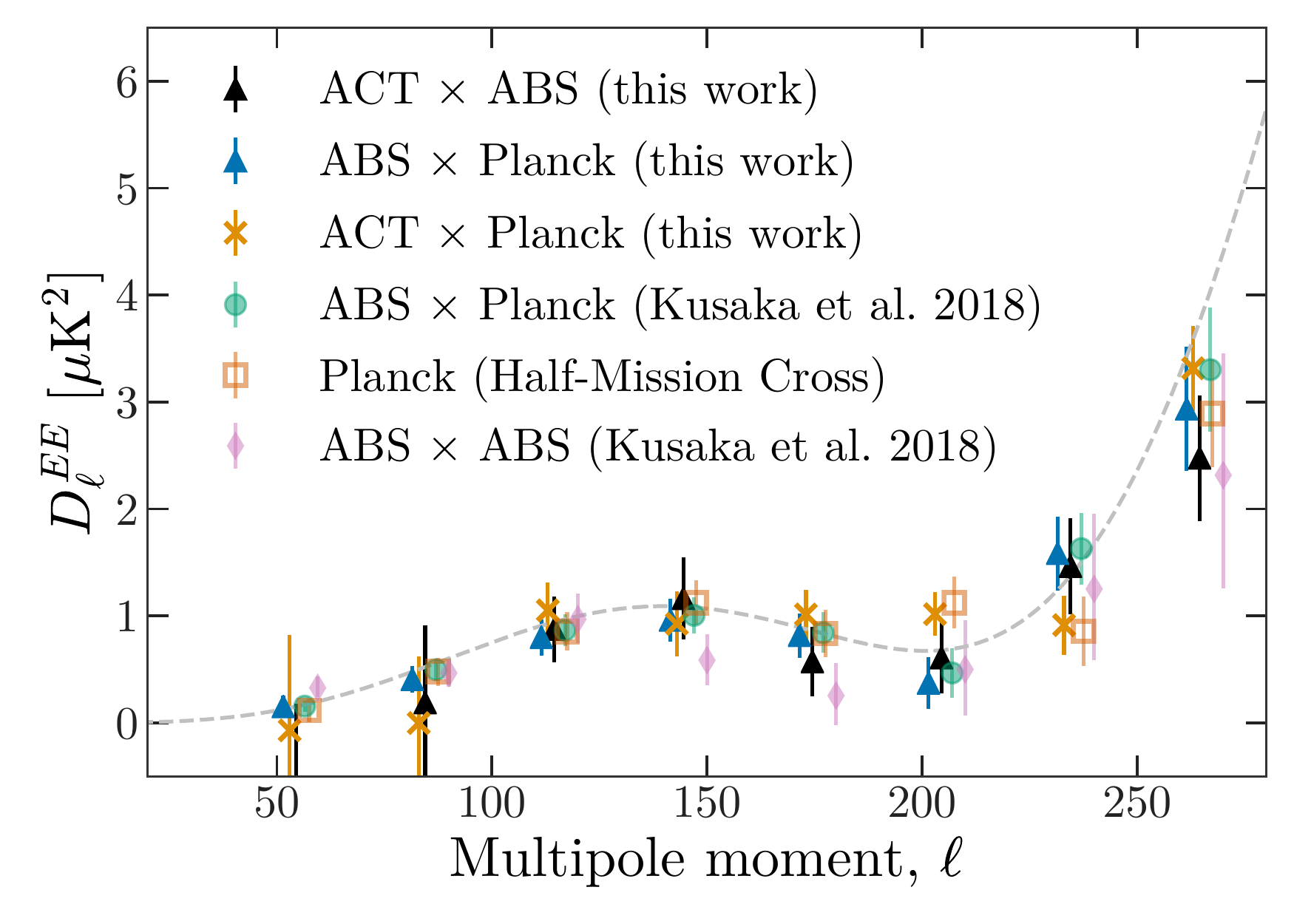}
    \caption{The same data as shown in Figure \ref{fig:EE}, but with an expanded vertical axis. Similarly, these spectra are from ACT, ABS, and Planck restricted to the ABS patch. A theoretical prediction for the spectrum from a $\Lambda$CDM model is shown in dashed gray, with parameters given in $\S$\ref{sec:analysis}.}
    \label{fig:EE_low}
\end{figure*}

Figures~\ref{fig:TE}, \ref{fig:EE}, and \ref{fig:EE_low} show the results along with the nominal $\Lambda$CDM model based on Planck, with comparisons to an earlier estimate of Planck $T \times$ ABS $E$ \citep{kusaka/etal:2018}. The data are given in Tables~\ref{tab:spectra1} and \ref{tab:spectra2}.
It is clear from the figures that all correlations are in agreement with Planck. We quantify the agreement with a simple $\chi^2$ statistic as given in Table~\ref{tab:chisq}. 
It is notable that the TE anticorrelation at $\ell=150$ can be seen with ground-based observations alone. 

There are a large number of cross spectra that should be consistent with zero. These too are reported in Table~\ref{tab:chisq}. There are no obvious trends in plots of the null spectra so they are omitted. We note that spectra related to B-modes in Planck in this $\ell$ range should be considered preliminary.


\begin{table}
\centering
\begin{tabular}{cccc} \toprule
   Instruments & Cross-Spectrum & $\chi^2$ (dof) & PTE \\ \hline
Planck $\times$ Planck & TB & 13.0 (13) & 0.45 \\
ACT $\times$ Planck & TB & 21.0 (13) & 0.07 \\
ACT $\times$ ABS & TB & 21.3 (13) & 0.07 \\
Planck $\times$ Planck & BT & 35.6 (13) & 0.00 \\
ABS $\times$ Planck & BT & 14.1 (13) & 0.37 \\
ACT $\times$ Planck & BT & 9.7 (13) & 0.72 \\
\hline
Planck $\times$ Planck & EB & 18.2 (13) & 0.15 \\
ABS $\times$ Planck & EB & 11.7 (13) & 0.55 \\
ACT $\times$ Planck & EB & 27.3 (13) & 0.01 \\
ACT $\times$ ABS & EB & 12.3 (13) & 0.50 \\
Planck $\times$ Planck & BE & 40.6 (13) & 0.00 \\
ABS $\times$ Planck & BE & 12.0 (13) & 0.53 \\
ACT $\times$ Planck & BE & 32.8 (13) & 0.00 \\
ACT $\times$ ABS & BE & 19.8 (13) & 0.10 \\
\hline
Planck $\times$ Planck & BB & 24.1 (13) & 0.03 \\
ABS $\times$ Planck & BB & 17.3 (13) & 0.19 \\
ACT $\times$ Planck & BB & 16.6 (13) & 0.22 \\
ACT $\times$ ABS & BB & 11.0 (13) & 0.61 \\
\hline
ACT $\times$ ABS & TE & 16.2 (12) & 0.182 \\
ACT $\times$ ABS & EE & 9.5 (12) & 0.660 \\
\end{tabular}
\caption{This table is filled with $\chi^2$ values for a model where the power spectrum is approximately zero (``null tests"). The TE and EE tests are performed for differences to Planck spectra on the same region, using covariance matrices computed without cosmic variance. The other nulls are performed relative to zero assumed signal, with covariance matrices that do include cosmic variance. \numbered{1}{We do not account for the error in estimated transfer function in ACT, as we expect the function to be highly correlated bin-to-bin. We anticipate that the final release of the ACT data will have a precisely measured transfer function, from injecting signal into the map-maker.} \numbered{8}{The foreground level in the ABS region is expected to be small, so we do not attempt any treatment of foregrounds in these nulls}. \numbered{9}{We also did not include Planck systematic effects such as polarization efficiency and beam leakage in these tests, which we expect would reduce the null tensions, as these results are not intended for precision cosmology.}
}
\label{tab:chisq}
\end{table}

Section $\S$\ref{sec:obs} illustrates how ABS and ACT were optimized differently and observe differently. For example, ABS uses a HWP. Other than the scientists in common and sharing the same site, they are completely independent experiments analyzed in completely different ways. 
From the cross correlation we conclude the following.

First, we confirm the ABS$\times$Planck correlation reported in \cite{kusaka/etal:2018} for both TE and EE, using a completely different pipeline for power spectra.
Any residual systematic effects in the ABS maps are subdominant to the signal. Not only does ABS pass a number of internal checks and constrain systematic effects to be significantly smaller than statistical ones, but it correlates well with Planck and now ACT.

Second, the output of different instruments observing in Chile can be directly compared. So far, analyses of ACT maps have been limited to $\ell>450$ because of the significant increase in noise from atmospheric fluctuations and potential ground contamination at lower $\ell$. This analysis shows that the polarization signal is still recoverable in the maps at least down to $\ell=100$. On-going efforts are aimed at further improving ACT's $\ell<450$ response. 

Third, the cross correlation reduces the uncertainties in the ABS experiment, especially at the upper end of the $\ell$-range. With better control of ACT's $\ell<450$ response, ABS could be improved over its full $\ell$ range.
Indeed, the ABS $\times$ Planck cross correlation improves ABS at all $\ell$, and ACT's noise per pixel is lower than Planck's. Such improvements open up new ways to constrain the low $\ell$ polarization and may aid in identifying primordial B-modes if they exist at measurable levels. These results are promising for future experiments such as the Simons Observatory \citep{SO:2019} which will have a suite of diffraction limited large (high, $\ell>500$) and small (low, $\ell<500$) aperture instruments.   



%

\section{Acknowledgements} 
This work was supported by the U.S. National Science Foundation through awards AST-0408698, AST-0965625, and AST-1440226, for the ACT project, as well as awards PHY-0355328, PHY-0855887 (both also for ABS), and PHY-1214379. The development of multichroic detectors and lenses was supported by NASA grants NNX13AE56G and NNX14AB58G. Detector research at NIST was supported by the NIST Innovations in Measurement Science program. ABS was also supported through NASA award NNX08AE03G, and the Wilkinson and Mishrahi funds. Funding was also provided by Princeton University, the University of Pennsylvania, and a Canada Foundation for Innovation (CFI) award to UBC. ACT, as did ABS, operates in the Parque Astron\'omico Atacama in northern Chile under the auspices of CONICYT. Computations were performed on the GPC and Niagara supercomputers at the SciNet HPC Consortium. SciNet is funded by the CFI under the auspices of Compute Canada, the Government of Ontario, the Ontario Research 
Fund---Research Excellence; and the University of Toronto. Flatiron Institute is supported by the Simons Foundation. ZL and JD are supported through NSF grant AST-1814971. Lastly, we thank Srinivasan Raghunathan for helpful comments on an earlier version of the paper.

\begin{table*}
\centering

\resizebox{\columnwidth}{!}{%

\begin{tabular}{ccccccccc}\toprule
$\ell$ & ACT T $\times$ ABS E & Error & Planck T $\times$ ABS E & Error & Planck T $\times$ ABS E  & Error & Planck $\times$ Planck  & Error \\
 &   &  & (this work) &  & (Kusaka et al. 2018)  &  \\
 \hline
54.5 & -7.3 & 18.4 & -1.9 & 5.2 & -3.8 & 2.5 & -4.5 & 4.3 \\
84.5 & -5.3 & 16.7 & -10.9 & 5.3 & -14.8 & 3.9 & -11.8 & 5.0 \\
114.5 & -34.7 & 17.9 & -30.9 & 7.0 & -27.5 & 5.5 & -18.6 & 6.5 \\
144.5 & -38.0 & 17.2 & -42.1 & 8.9 & -46.4 & 7.0 & -43.4 & 7.9 \\
174.5 & -24.7 & 17.2 & -18.7 & 10.0 & -25.3 & 9.1 & -39.3 & 8.6 \\
204.5 & -26.6 & 18.1 & -8.7 & 11.5 & -14.3 & 10.1 & 6.6 & 8.9 \\
234.5 & 15.2 & 20.6 & 29.8 & 13.9 & 42.5 & 11.5 & 31.9 & 9.9 \\
264.5 & 79.4 & 19.9 & 75.8 & 16.5 & 69.0 & 13.3 & 69.1 & 11.6 \\
294.5 & 87.6 & 23.4 & 77.0 & 18.9 & 76.4 & 14.7 & 111.6 & 13.1 \\
324.5 & 95.2 & 25.0 & 114.9 & 20.0 & 117.9 & 16.6 & 109.3 & 13.2 \\
354.5 & 42.7 & 25.5 & 51.8 & 20.2 & 40.3 & 16.0 & 82.3 & 12.0 \\
384.5 & 25.8 & 26.7 & 32.1 & 20.6 & 24.0 & 17.0 & 39.2 & 10.6 \\
414.5 & -65.4 & 32.8 & -52.1 & 24.1 & -51.2 & 20.7 & -48.6 & 10.0 \\
\hline
\end{tabular}
}
\caption{Bandpowers from the TE cross-correlations. Units are in $\mu$K$^2$. The $\ell$ column refers to the band center.}
\label{tab:spectra1}
\end{table*}

\bigskip

\begin{table*}
\centering
\begin{tabular}{cccccccc}\toprule
$\ell$ & ACT $\times$ ABS &  Error & ABS $\times$ Planck &  \numbered{2}{Error}  & Planck  &   Error \\
 &  &  & (this work) & & (Half-Mission Cross) &   \\
 \hline
54.5 & -0.8 & 1.6 & 0.2 & 0.1 & -0.1 & 1.5 \\
84.5 & 0.2 & 0.6 & 0.4 & 0.1 & -0.0 & 0.5 \\
114.5 & 0.9 & 0.5 & 0.8 & 0.2 & 1.1 & 0.4 \\
144.5 & 1.2 & 0.4 & 1.0 & 0.2 & 0.9 & 0.3 \\
174.5 & 0.6 & 0.3 & 0.8 & 0.2 & 1.0 & 0.2 \\
204.5 & 0.6 & 0.3 & 0.4 & 0.2 & 1.0 & 0.2 \\
234.5 & 1.5 & 0.4 & 1.6 & 0.3 & 0.9 & 0.3 \\
264.5 & 2.5 & 0.6 & 2.9 & 0.6 & 3.3 & 0.4 \\
294.5 & 8.0 & 1.1 & 6.8 & 1.0 & 9.0 & 0.7 \\
324.5 & 13.3 & 1.6 & 12.9 & 1.5 & 13.4 & 1.1 \\
354.5 & 18.2 & 2.3 & 15.7 & 2.0 & 19.9 & 1.4 \\
384.5 & 21.8 & 2.8 & 19.0 & 2.5 & 23.2 & 1.6 \\
414.5 & 26.3 & 3.8 & 22.0 & 3.1 & 28.6 & 1.7 \\
\hline
\end{tabular}
\caption{Bandpowers from the EE correlations. Units are in $\mu$K$^2$. The $\ell$ column refers to the band center. \numbered{2}{Bandpowers are in $D_{\ell} = \ell (\ell + 1) C_{\ell} / 2\pi$}.}
\label{tab:spectra2}
\end{table*}

\bibliographystyle{wmap}
\bibliography{apj-jour, absact_refs}
  
\end{document}